\documentclass[prb,aps,twocolumn,showpacs,home,floats]{revtex4}

\usepackage{graphics}
\begin{document}

\title{  {\rm\small\hfill (submitted to Appl. Phys. A})\\
Oxide formation at the surface of late \boldmath$4d$ transition metals:\\
Insights from {\em first-principles atomistic thermodynamics}}

\author{Karsten Reuter${}^{1,2}$ and Matthias Scheffler${}^{1}$}

\affiliation{${}^1$ Fritz-Haber-Institut der Max-Planck-Gesellschaft, Faradayweg
4-6, D-14195 Berlin-Dahlem, Germany}
\affiliation{${}^2$ FOM Institute for Atomic and Molecular Physics, Kruislaan 407,
1098 SJ Amsterdam (The Netherlands)}

\date{Received: 24 March 2003}

\begin{abstract}
Using density-functional theory we assess the stability of bulk and 
surface oxides of the late $4d$ transition metals in a ``constrained equilibrium'' 
with a gas phase formed of O${}_2$ and CO. While the stability range of 
the most stable bulk oxide extends for ruthenium well into gas phase 
conditions representative of technological CO oxidation catalysis, this 
is progressively less so for the $4d$ metals to its right in the 
periodic system. Surface oxides could nevertheless still be stable under 
such conditions. These thermodynamic considerations are discussed in the light 
of recent experiments, emphasizing the role of (surface) oxides as the 
active phase of model catalysts formed from these metals. 
\end{abstract}



\maketitle

\section{Introduction} 

Rhodium and palladium exhibit a high reactivity for the CO oxidation
reaction, which has led to their widespread use in catalytic car
exhaust converters \cite{taylor93}. The degree of oxidation of these 
materials in the reactive environment has been a frequent object of 
speculation in the experimental catalysis literature (see e.g. refs. 
\onlinecite{oh83,kellog85,peden88,turner81,bondzie96,ziauddin97,veser99}),
with oxide formation at the surface of the metal particles potentially
having either beneficial or detrimental effects on the overall
activity. Lacking microscopic {\em in-situ} data to really resolve this
issue a possibly beneficial role has e.g. been suspected for PdO at Pd 
\cite{ziauddin97}, whereas Rhodium oxides have predominantly been 
viewed as poisoning the catalytic reaction \cite{oh83,kellog85,peden88}. 

While the atomic-scale techniques of Ultra High Vacuum (UHV) surface-science 
would in principle be able to provide additional insight into this 
subject matter, formation of oxides in particular of the more 
noble metals was often neglected in corresponding works, partly due 
to the fact that these oxides form only at rather high oxygen pressures 
and elevated temperatures. This emphasis on metallic substrates has
changed over the last years, bringing oxide surfaces now also on the
agenda of surface-science studies attempting to bridge the pressure gap 
between UHV and technological oxidation catalysis \cite{stampfl02}. 
Concomitantly, on Ru(0001) \cite{over00,kim01b}, Pd(100) 
\cite{hendriksen03a,hendriksen03b}, and Ag(111) \cite{carlisle00b,li03}
oxygen-rich environmental conditions were reported to lead to oxide 
(or ``surface oxide'') formation, in all cases connected with a
significant increase in catalytic activity.

This new focus on the effect of the environment on surface structure
and reactivity is intertwined with corresponding theoretical efforts.
In particular the concept of {\em first-principles atomistic thermodynamics} 
\cite{weinert86,scheffler88,kaxiras87,qian88} has recently proven
to be most valuable for the study of oxide surfaces (see e.g. refs. 
\onlinecite{wang98,wang00,reuter02a,reuter03a,reuter03b}). In this
branch of thermodynamics density-functional theory (DFT) is employed to 
compute free energies, in order to identify the lowest-energy atomic
structure for a given condition of thermodynamic reservoirs representing 
the surrounding gas phase, and possibly combining this with constraints
to facilitate the analysis \cite{reuter03a,reuter03b}. In the following 
we will use and describe this formalism to discuss the stability of bulk 
and surface oxides of the late $4d$ transition metals (TMs) Ru, Rh, Pd, 
and Ag. The results of these very basic thermodynamic considerations are 
found to nicely embrace the aforedescribed existing experimental data, e.g. 
pointing at the relevance of thin surface oxide layers on Pd under the 
conditions of high-pressure CO oxidation catalysis.

\section{``Constrained thermodynamics''} 

The question we like to ask in this context is, given a certain 
gas phase environment composed of O${}_2$ and CO, does a bulk or surface
oxide of the late $4d$ TMs form a stable phase? In a thermodynamic
description we then resort to some form of equilibrium that exists between 
various components $i$ in the system, all of which are present in sufficient 
quantities. Under conditions of constant temperature, $T$, and pressures, 
$\{p_i\}$, this allows to discuss relations between the respective reservoirs, 
characterized by their chemical potentials $\mu_i(T,p_i)$ or 
Gibbs free energies $g_i(T,p_i)$. For a gas phase composed of CO and O${}_2$ 
a forthright assumption of full equilibrium does, however, not yield a description 
suitable for our interest in oxidation catalysis: From energy considerations 
alone, CO${}_2$ would then result as the most stable gas phase molecule for 
almost all temperature and pressure conditions. On the other hand, because of the 
large free energy barrier for the gas phase reaction CO + 1/2 O${}_2 \rightarrow$ CO${}_2$, 
such a full equilibrium will not be attained on time scales relevant to us. Instead, 
we may ignore CO${}_2$ formation in the gas phase and consider the presence of 
two separate, independent reservoirs for O$_2$ and CO in the environment. While 
both components are therefore not in equilibrium with each other, each one is 
individually assumed in equilibrium with either the metal (M) or the oxide 
(${\rm M_xO_y}$) phase.

As a consequence the chemical potentials of both oxygen and CO in the whole
system are then determined by the surrounding gas phase reservoirs, i.e. their 
temperature and pressure dependence is given by 
\cite{reuter02a,reuter03b}
\begin{equation}
\mu_{\rm O}(T,p_{\rm O_2}) = 
\frac{1}{2}\left[ E^{\rm total}_{\rm O_2}
+ \tilde{\mu}_{\rm O_2}(T,p^{0}) 
+ k_{\rm B} T ln \left( \frac{p_{\rm O_2}}{p^{0}} \right) \right],
\label{eq4}
\end{equation}
and
\begin{equation}
\mu_{\rm CO}(T,p_{\rm CO}) =
E^{\rm total}_{\rm CO}
 + \tilde{\mu}_{\rm CO}(T,p^{0})
+ k_{\rm B} T ln \left( \frac{p_{\rm CO}}{p^{0}} \right),
\label{eq5}
\end{equation}
where the temperature dependence of $\tilde{\mu}_{{\rm O}{_2}}(T,p^{0})$
and $\tilde{\mu}_{\rm CO}(T,p^{0})$ includes the contributions from
vibrations and rotations of the molecules, as well as the ideal gas
entropy at $p^{0} = 1$\,atmosphere. The latter quantities can be computed
from first principles, yielding results that are at the temperature range 
of interest to us virtually indistinguishable from the experimental values 
listed in thermodynamic tables \cite{reuter03b,JANAF}. Via eqs. 
(\ref{eq4}) and (\ref{eq5}) our results obtained as a function of the 
chemical potentials, i.e. in $(\mu_{\rm O}, \mu_{\rm CO})$-space, can then 
be converted into pressure scales at any specific temperature. Roughly 
representing the temperature range relevant to our study, this will be 
illustrated with specific scales at $T=300$\,K and $T=600$\,K in all figures 
below.

In the resulting ``constrained equilibrium'' situation \cite{reuter03a,reuter03b} 
the only pathway to CO${}_2$ formation is due to a reduction of the oxide or due 
to the catalytic CO oxidation at the substrate surface. In a flow reactor
this constant build-up of CO${}_2$ is counteracted by the continuous stream
(and thus removal) of gases over the catalyst surface, yielding a CO${}_2$ 
concentration gradient under stationary operation conditions. Although formed 
CO${}_2$ molecules may in principle readsorb at the surface (dissociatively or as a molecule), 
the probability of this event is very low compared to the much more frequent
O${}_2$ and CO (re)adsorption. The surface is therefore unlikely to equilibrate
with the surrounding CO${}_2$ gas phase, and a proper description would have to 
involve a dynamical treatment of the CO${}_2$ gas flow close to the catalyst 
surface. Such treatment will be published elsewhere \cite{reuter03c}. Here we
use a very simple estimate for the CO${}_2$ free energy, namely the internal 
energy of a free CO${}_2$ molecule. The error introduced by this estimate will 
vary from material to material, but in our preliminary kinetic Monte Carlo 
simulations for RuO${}_2$(110) we find it to be small.\cite{reuter03c}

Although crude, this approximation is appropriate for the present trend study 
aiming only at a first assessment of oxide stability, and an uncertainty of some 
tenths of eV does not change the nature of our conclusions. In this respect we
stress that the whole concept of a ``constrained equilibrium'' is not
designed to yield precise answers to the microscopic state and reactivity
of the catalyst surface. For that, the very dynamic behavior must be
modeled by statistical mechanics. Yet, with atomistic thermodynamics one
can trace out the large scale behavior and identify those conditions where such
a more refined treatment is necessary.

\section{Bulk oxide stability} 

If there is a pure oxygen environment, the condition for the stability of 
the bulk oxide is
\begin{equation}
g^{\rm bulk}_{{\rm M}_x{\rm O}_y} \; < \; 
x \; g^{\rm bulk}_{\rm M} + y \; \mu_{\rm O},
\end{equation}
i.e.
\begin{equation}
\Delta \mu_{\rm O} > \frac{1}{y} \left[ g^{\rm bulk}_{{\rm M}_x{\rm O}_y} 
- x \; g^{\rm bulk}_{\rm M} - \frac{y}{2} E^{\rm total}_{\rm O_2} \right].
\label{O2-condition}
\end{equation}
Here $\Delta \mu_{\rm O}$ is defined as $\mu_{\rm O} - (1/2) 
E^{\rm total}_{\rm O_2}$. For $T=0$ K the bracket on the right hand side 
equals half of the low temperature limit of the heat of formation, 
$H_f(T=0\,{\rm K}, p=0)$. Even for higher temperatures this will not change 
substantially as for bulk phases the $(T,p)$-dependence of their Gibbs 
free energies is rather small, i.e. the right hand side will still be
rather close to the $H_f(T=0\,{\rm K}, p=0)$-value and not follow the
pronounced variation of $H_f(T, p)$ (which is primarily due to the
$(T,p)$-dependence of the $\mu_{\rm O_2}$-term, while only $E^{\rm total}_{\rm O_2}$
enters the right hand side of eq. (\ref{O2-condition})). We therefore
replace the bracket in eq. (\ref{O2-condition}) by $(1/y)H_f(T=0\,{\rm K})$ 
and arrive at our first stability condition
\begin{equation}
\Delta \mu_{\rm O} \stackrel{\textstyle>}{\sim} \frac{1}{y} H_f(T=0\,{\rm K}).
\label{conditionI}
\end{equation}

Additionally, the oxide can also be destroyed (reduced) by carbon monoxide.
In a pure CO environment, the stability condition for the oxide is
\begin{equation}
g^{\rm bulk}_{{\rm M}_x{\rm O}_y} + y \; \mu_{\rm CO} \; < \;
x \; g^{\rm bulk}_{\rm M} + y \; \mu_{\rm CO_2}.
\label{OCOstab}
\end{equation}
Following the discussion above we will crudely approximate $\mu_{\rm CO_2}$
in the present study by the internal energy of a free CO${}_2$ molecule.
In a similar fashion as for the pure oxygen environment one can then simplify
the stability condition to 
\begin{equation}
\Delta \mu_{\rm CO} \stackrel{\textstyle<}{\sim} 
- \frac{1}{y} \; H_f(T=0\,{\rm K}) \; + \; \Delta E^{\rm mol.},
\label{pureCOstab}
\end{equation}
introducing $\Delta \mu_{\rm CO} = \mu_{\rm CO} - E^{\rm total}_{\rm CO}$ and
using the short hand $ \Delta E^{\rm mol.} = (E^{\rm bind}_{\rm CO_2} -
E^{\rm bind}_{\rm CO} - 1/2 E^{\rm bind}_{\rm O_2}) $ for the difference
in binding energies of the three gas phase molecules. As described in
detail in a previous publication \cite{reuter03b} our DFT computations
including zero-point vibrations give $\Delta E^{\rm mol.} = -3.2$\,eV, which
agrees well with the experimental value of $-2.93$\,eV \cite{JANAF}.
Although state-of-the-art DFT total energies of the O$_2$ molecule are 
subject to a noticeable error, some cancelation apparently occurs for 
the composite quantity $\Delta E^{\rm mol.}$, rendering it rather unimportant 
on the scale of the present study whether the experimental or theoretical 
value for $\Delta E^{\rm mol.}$ is used.

\begin{table}
\caption{\label{tableI}
Low-temperature limit of the heat of formation, $\Delta H_f(T=0\,{\rm K})$,
of the most stable bulk oxide of the four late $4d$ transition metals from 
DFT and from experiment. The DFT values do not include the zero-point energies
of the solid phases.}
\begin{tabular}{l @{\quad\quad} | @{\quad\quad}l @{\quad\quad} l}
\                & Theory                        &  Experiment            \\ \hline
${\rm RuO_2}$   & $-3.4$\,eV \cite{reuter02a}   &  $-3.19$\,eV \cite{CRC}  \\
${\rm Rh_2O_3}$ & $-3.8$\,eV \cite{grillo03}    &  $-3.57$\,eV \cite{CRC}  \\
${\rm PdO}$     & $-0.9$\,eV \cite{rogal03}     &  $-0.88$\,eV \cite{CRC}  \\
${\rm Ag_2O}$   & $-0.2$\,eV \cite{li03}        &  $-0.34$\,eV \cite{CRC}  \\
\end{tabular}
\end{table}

Finally, if O${}_2$ and CO are both present in the gas phase in
``constrained equilibrium'' with the oxide, then the general stability
condition is simply obtained by combining eqs. (\ref{conditionI}) and
(\ref{pureCOstab}) for the separate cases:
\begin{equation}
\Delta \mu_{\rm CO} - \Delta \mu_{\rm O} \stackrel{\textstyle<}{\sim}
- \frac{2}{y} H_f(T=0\,{\rm K}) \; + \; \Delta E^{\rm mol.},
\label{conditionII}
\end{equation}
while eq. (\ref{conditionI}) still applies for very low CO concentrations.
Hence, in both conditions for the stability of the bulk oxide, eqs. 
(\ref{conditionI}) and (\ref{conditionII}), the only quantity left to determine 
is $\Delta H_f(T=0\,{\rm K})$. This value has been computed in previous works
\cite{li03,reuter02a,grillo03,rogal03}
and we content ourselves here with simply listing the DFT values for the most
stable oxides of the four late $4d$ TMs in Table \ref{tableI}, again comparing 
with the low-temperature limit of the experimental heat of formation
\cite{CRC}. Keeping 
the sizable error in the O${}_2$ DFT total energy in mind, we notice in all cases 
a fortuitous error cancelation, leading again to rather small differences 
when evaluating the stability conditions with either experimental or theoretical 
numbers. Only in the case of silver oxide has $\Delta H_f(T=0\,{\rm K})$ become
so small that this error becomes somewhat disturbing (in particular for
the stability condition in eq. (\ref{conditionII})). More refined treatments
for approximating the stability conditions, e.g. including entropic contributions 
to $\Delta H_f(T,p)$, can be devised \cite{li03}, but for our present purposes
it suffices to see that bulk Ag${}_2$O has such a low stability that it will
not play a role in our further discussion.

\begin{figure}
\scalebox{0.31}{\includegraphics{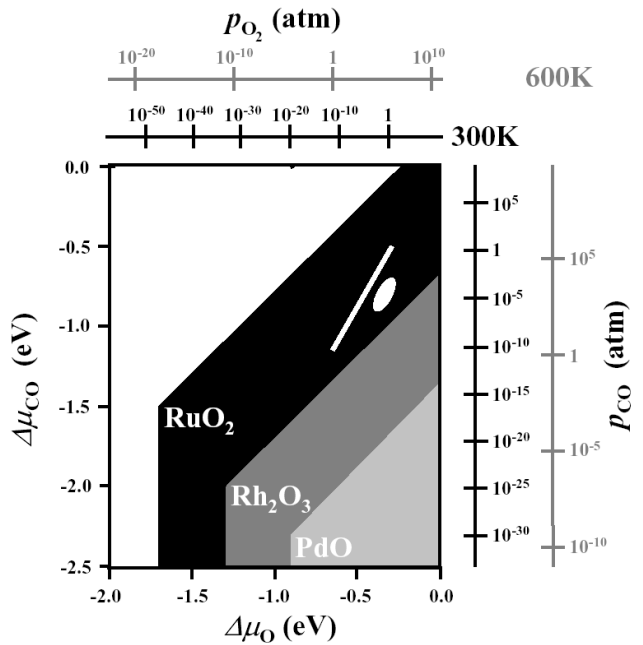}}
\caption{Stability regions of bulk oxides of the late $4d$ transition
metals in $(\Delta \mu_{\rm O}, \Delta \mu_{\rm CO})$-space, on the
basis of the DFT data listed in Table I. Additionally, pressure scales are 
drawn for $T = 300$\,K and $T = 600$\,K. The white line indicates conditions 
corresponding to $p_{\rm O} = p_{\rm CO} = 1$\,bar and 300\,K$\le T \le$ 600\,K,
and environments relevant for CO oxidation catalysis over these metals would
correspond to the near vicinity of this line. More specifically, the small
white area indicates roughly the $(T,p)$-conditions employed in a recent experimental 
study by Hendriksen and Frenken (see text) \cite{hendriksen03b,hendriksen03c}. 
The stability of Ag${}_2$O is so low, that is does not appear any more within the
limits considered here. We stress that the rough treatment of the CO${}_2$
free energy translates into an uncertainty in the position of the diagonal
lines (which could be shifted up or down by some tenths of an eV).}
\label{fig1}
\end{figure}

The derived $(T,p)$-ranges for stability of the $4d$ bulk oxides are shown in 
Fig. \ref{fig1}, exhibiting a pronounced decrease from RuO${}_2$ towards the
oxides of the metals to its right in the periodic system. We stress that the
aforediscussed crude treatment of the CO${}_2$ free energy translates into
rather large error bars on the diagonal lines running from bottom left to top
right in the figure (i.e. the limit due to the stability condition of eq. 
(\ref{conditionII})). The uncertainty in $\mu_{\rm CO_2}$ translates directly
into the height of these lines, which with the present error estimate could
therefore be shifted up or down by some tenths of an eV.

\section{Surface oxide stability}

Defering the discussion of the derived bulk oxide stability to the end of this 
paper, we proceed to analyze in the stability context a second, only recently 
emphasized aspect of metal oxidation, namely the formation of thin surface oxides. 
While traditionally such films were conceived as closely-related thin versions 
of the corresponding bulk oxides, recent atomic scale characterizations of 
initial few-atom thick oxide overlayers especially on Pd and Ag surfaces revealed 
structures that had only little resemblance to their bulk counterparts, and/or 
were to a large degree influenced by a strong coupling to the underlying metal 
substrate \cite{carlisle00b,lundgren02,todorova03}. Due to this coupling
and structures particularly suited for layered configurations, one may expect 
the stability range for such surface oxides to exceed that of the hitherto
discussed bulk oxides. 

To determine the range of $(T,p)$-conditions in which a surface oxide would
represent the thermodynamically most stable state, we follow the approach of Li, 
Stampfl and Scheffler \cite{li03}, i.e. we evaluate the Gibbs free energy of 
adsorption and compare it to other possible states of the system like on- or 
sub-surface oxygen phases or the bulk oxide. For the case of a pure O${}_2$ 
environment and using the same approximations as discussed in the last section, we 
may write this Gibbs free energy of adsorption as
\begin{eqnarray}
\label{surfoxstab} \nonumber
\lefteqn{\Delta G(\Delta \mu_{\rm O}) \approx} && \\ \nonumber
&\approx& - \frac{1}{A} \left( E^{\rm total}_{\rm O@M}
- E^{\rm total}_{\rm M} - N_{\rm O} \left( \frac{1}{2} E^{\rm total}_{\rm O_2} +
\Delta \mu_{\rm O} \right) \right) \\
&=& \frac{N_{\rm O}}{A} \left( E^{\rm bind}_{O@M} + \Delta \mu_{\rm O} \right).
\end{eqnarray}
Here, $E^{\rm total}_{\rm O@M}$ and $E^{\rm total}_{\rm M}$ are the total energies
of the surface with and without oxygen coverage of $N_{\rm O}$ oxygen atoms per
surface area $A$. In
the second line, we have identified the first terms in the brackets with the
average binding energy of oxygen in the particular surface configuration and with
respect to 1/2 $E^{\rm total}_{\rm O_2}$. 

Obviously, the higher the oxygen content of a considered surface structure, the steeper 
its $\Delta G(\Delta \mu_{\rm O})$ will decrease with increasing chemical potential. In 
the limiting case of an infinitely thick bulk oxide on top of the metal substrate, this 
will result in a vertical line that crosses the zero-axis at the
stability condition for the bulk oxide in eq. (\ref{conditionI}). For any
higher $\Delta \mu_{\rm O}$ the bulk oxide will be the stable phase, but the interesting
question is to see whether there are lower $\Delta \mu_{\rm O}$ than this limit, for
which the lines of a surface oxide structure turn out lower than e.g. on-surface 
adsorption or the clean surface corresponding to $\Delta G(\Delta \mu_{\rm O})=0$.

\begin{figure}
\scalebox{0.39}{\includegraphics{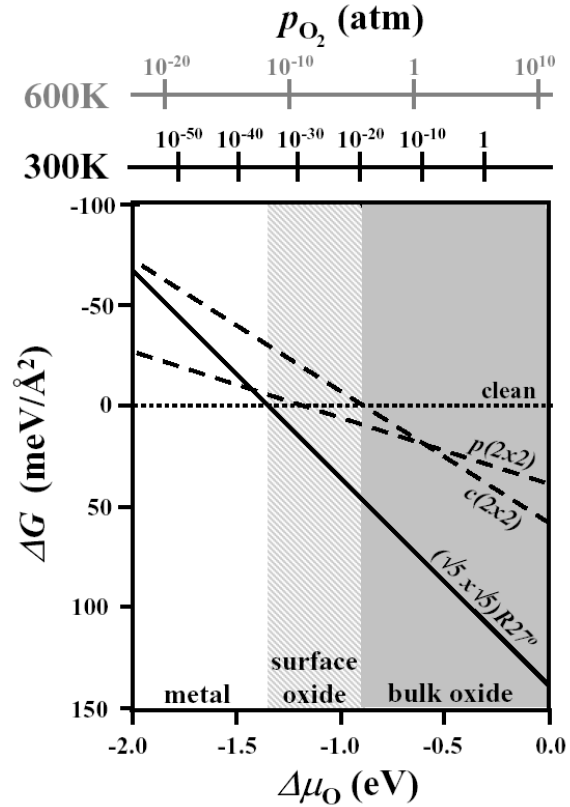}}
\caption{Computed Gibbs free energy of adsorption for the $p(2 \times 2)$ and 
$c(2 \times 2)$ on-surface adsorption phase, as well as for the $(\sqrt{5} 
\times \sqrt{5})R27^o$-O surface oxide on Pd(100). The surface unit cell of 
Pd(100) is 7.8\,{\AA}${}^2$. The stability range of
the surface oxide extends well beyond that of the bulk oxide, given by
$\Delta \mu_{\rm O} \stackrel{\textstyle>}{\sim} -0.9$\,eV, cf. eq.
(\ref{conditionI}). The dependence on $\Delta \mu_{\rm O}$ is again
translated into pressure scales at $T=300$\,K and $T=600$\,K for clarity. 
In the bottom of the figure, the ``material type'' which is stable in the 
corresponding range of O chemical potential is listed and indicated by the 
shaded regions. The binding energies used to construct this graph are
taken from refs. \onlinecite{todorova03,todorova03b}.}
\label{fig2}
\end{figure}

We exemplify this ansatz by first analyzing the stability of a surface oxide on Pd(100).
In addition to the well known $p(2 \times 2)$ and $c(2 \times 2)$ ordered adlayers 
with O in fcc sites \cite{kolthoff96,zheng02}, an ensuing $(\sqrt{5} \times \sqrt{5})R27^o$-O
surface oxide phase on this surface was recently characterized as a rumpled, but 
commensurate PdO(101) film with a strong coupling to the underlying Pd(100) substrate 
\cite{todorova03,todorova03b}. Evaluating the reported binding energies in this study with
eq. (\ref{surfoxstab}) we obtain the results displayed in Fig. \ref{fig2}. Interestingly,
the $\Delta \mu_{\rm O}$-range where this surface oxide exhibits the lowest
Gibbs free energy of adsorption not only exceeds the one of the aforediscussed
PdO bulk oxide, but also the range where the on-surface adsorption phases are
more stable than the clean Pd(100) surface. In other words, the latter on-surface 
phases never correspond to a thermodynamically stable phase, and their frequent
observation in UHV experiments \cite{kolthoff96,zheng02} appears to be a mere outcome of
the limited O supply offered, as well as of kinetic barriers e.g. for O penetration
at the low temperatures employed (UHV experiments are typically performed by depositing
a finite number of adatoms, rather than by maintaining a given gas pressure). 

\begin{figure}
\scalebox{0.39}{\includegraphics{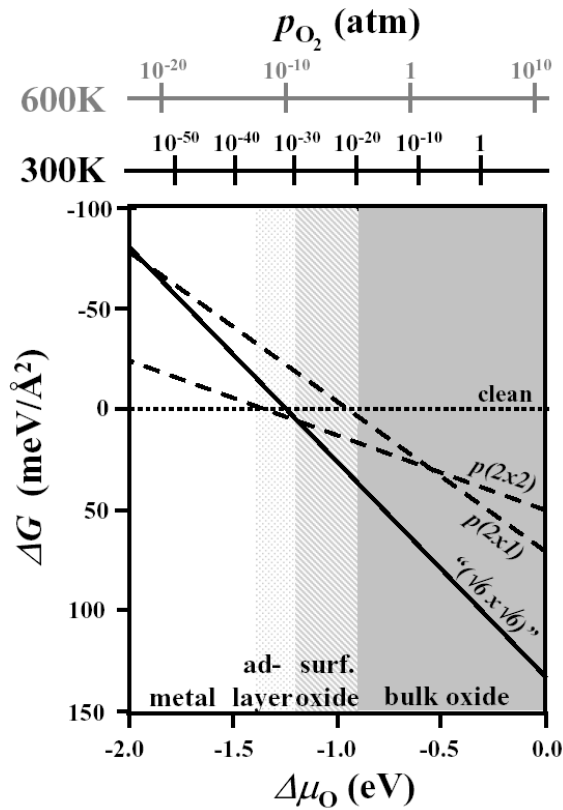}}
\caption{Computed Gibbs free energy of adsorption for the $p(2 \times 2)$ and 
hypothetical $p(2 \times 1)$ on-surface adsorption phase, as well as for the 
so-called $(\sqrt{6} \times \sqrt{6})$-O surface oxide on Pd(111). The surface 
unit cell of Pd(111) is 6.7\,{\AA}${}^2$. Again, the stability range of
the surface oxide extends beyond that of the bulk oxide, cf. Fig. \ref{fig2},
but this time there is also a finite range in which an adlayer forms the
thermodynamically most stable phase. The binding energies used to construct this 
graph are taken from ref. \onlinecite{lundgren02}.}
\label{fig3}
\end{figure}

It is interesting to compare these findings with the equivalent situation on
another Pd surface orientation. On Pd(111), a surface oxide not resembling
the bulk-structure of PdO at all was recently reported as the product of high
oxygen exposure and following the well-studied $p(2 \times 2)$ on-surface
adsorption phase \cite{lundgren02}. Using the DFT binding energies provided
in this study to evaluate eq. (\ref{surfoxstab}), we again obtain a range of
$(T,p)$-conditions outside of the bulk oxide stability range, for which
this surface oxide phase is most stable as shown in Fig. \ref{fig3}. This time, 
however, this range extends only within -1.2\,eV $\stackrel{\textstyle>}{\sim} 
\Delta \mu_{\rm O} \stackrel{\textstyle>}{\sim}-0.9$\,eV, below which first the 
on-surface adsorption phase and then the clean Pd(111) surface becomes more stable.
If we compare this to the results shown in Fig. \ref{fig2}, there are thus
environmental conditions, in which a surface oxide may already be 
thermodynamically stable on Pd(100), while on-surface adsorption still
prevails on Pd(111). Knowing that these two surfaces form the predominant
surface area of Pd nanoparticles \cite{szanyi93}, this may have profound
consequences on the oxidation behavior of the latter and we will return to
this point below.

While a similar situation with the commensurable $p(4 \times 4)$ surface oxide 
\cite{carlisle00b} slightly exceeding the stability range of bulk Ag${}_2$O was 
also found for Ag(111) \cite{li03}, no $(T,p)$-range most favorable for the various
intermediate precursors suggested in the oxidation pathway from Ru(0001)
to RuO${}_2$(110) is obtained when evaluating the DFT binding energies reported
in the corresponding refs. \onlinecite{reuter02b,reuter02c}. Tentatively, we 
therefore expect surface oxides to play a thermodynamic role more for the $4d$ 
TMs towards the right of the periodic system, where the decreasing thermal stability 
of the bulk oxides and the lowered bulk modulus of oxide and metal phase enhances 
the influence of oxide-metal coupling,  and thus the tendency to form 
commensurable, even non-bulk like surface oxide configurations. It is
interesting to notice that the extended stability range of the surface oxide phase 
discussed on Pd(100), cf. Fig. \ref{fig2}, corresponds more or less to
the range exhibited by bulk Rh${}_2$O${}_3$ in Fig. \ref{fig1}, i.e. it
comes now rather close to the schematically indicated range of 
$(T,p_{\rm O},p_{\rm CO})$-conditions representative of technological CO
oxidation catalysis.

\section{Implications for the catalytic CO oxidation}

Due to the particular construction of our ``constrained equilibrium'' approach
we could expect the results summarized in Figs. \ref{fig1} to \ref{fig3} to give 
a rough reflection of the state of the catalyst during steady-state CO oxidation,
as long as kinetic effects both in the formation of the (surface) oxides
themselves, as well as in the ongoing reaction on the surface (e.g. possibly
consuming oxygen more rapidly than it is replenished) are negligible
\cite{reuter03c}. Although
technological catalysis typically operates well above room temperature,
such effects can never be ruled out in general. Clearly, a comprehensive
understanding of the full catalytic activity and the microscopic state of
the catalyst (i.e. its very dynamic behavior) requires modeling by statistical 
mechanics, and the results of thermodynamic considerations must consequently
be treated with great care. Still, we believe that our results may be employed 
for a first qualitative assessment of oxide formation on the late $4d$ TMs,
in particular if supplemented by some already known kinetic constraints.

In this regard we first notice the rather large stability range of bulk
RuO${}_2$, extending well over the whole range of environmental conditions
relevant for technological CO oxidation catalysis, cf. Fig. \ref{fig1}. 
Rather high temperatures of the order of $T \sim 600$\,K are required
to form this bulk oxide \cite{over00,kim01b}, which would explain the 
recently reported lack of oxide formation at Ru(0001) even under high pressure 
conditions at temperatures around $T \sim 400$\,K \cite{hendriksen03b,hendriksen03c}. 
Yet, once this bulk oxide is formed, our thermodynamic results imply that the 
catalytic activity of Ru catalysts might in reality be due to oxides, in line 
with the results of a number of recent experimental studies 
\cite{stampfl02,over00,kim01b}.

For Rh the stability range of its bulk oxide has already decreased significantly.
Even if the catalyst was in a situation close to thermodynamic equilibrium,
our results suggest that either metal or bulk oxide could prevail at
technologically relevant environmental conditions, depending on the exact 
partial pressures and temperature in the gas phase, i.e. depending on
whether one is below or above the stability line in Fig. \ref{fig1} (and recall that 
the present uncertainty in the location of this line is some tenths of eV). Even 
more intriguing, oscillations between the two phases could result as a
consequence of fluctuations, when operating under environmental conditions 
very close to the stability line of Rh${}_2$O${}_3$. Similar considerations might
also be possible for Pd, yet this time not concerning transitions to bulk PdO 
(which is already far too unstable), but to thin oxide films limited to the 
very surface region - as mentioned in section IV the stability of the surface
oxide on Pd(100) extends over a similar range to that of bulk Rh${}_2$O${}_3$
in Fig. \ref{fig1}. In this respect we have also roughly marked in Fig. \ref{fig1} 
the $(T,p)$-conditions employed in a recent experimental study on Pd(100), where
a substantial roughening of the surface with presumed constant formation and 
reduction of ultra-thin oxide domains during the CO oxidation reaction was 
reported \cite{hendriksen03a,hendriksen03b}. The closeness of these experimental 
environmental conditions with our rough instability estimate for the surface 
oxide on Pd(100) is startling, in particular when further recalling the 
frequent observation of spatio-temporal pattern formation during the CO 
reaction at other Pd surfaces like Pd(110), which was there also connected to 
reversible oxide formation and reduction \cite{bondzie96}. 

At this stage it is worth digressing on the fact that such thin oxide films with 
possibly even fluctuating conditions at the catalyst surface are tough to describe (even 
conceptually) within the traditional language of generic reaction mechanisms (see e.g.
ref. \onlinecite{masel96} for a review). For metal surfaces mostly a Langmuir-Hinshelwood 
(LH) type behavior is discussed, where both reactants adsorb at the surface 
first and react thereafter. For the case of oxidation reactions on oxide
surfaces on the other hand, Mars and van Krevelen (MvK) \cite{mars54} suggested 
that the underlying substrate itself might become continually reduced and (re-)oxidized
in the on-going reaction, i.e. that oxygen from the oxide lattice is consumed
in the reaction and then replenished from the gas phase. Within the framework
of macroscopic rate equations, where this approach was made and has ever since
been very popular (in the oxide literature), this translates to relaxing the constraint 
of a fixed number of active surface sites underlying the LH kinetics. The resulting 
MvK rate equation is then similar, but not identical to the one derived for a LH 
reaction mechanism, although in actual practice most data can be fitted equally 
well with both rate equations \cite{masel96}. 

This only vague distinction between both mechanisms becomes even more blurred when 
starting to think microscopically and particularly when considering what is now 
called a surface oxide (a 1-2 layer thick oxide-like film). For the case of a 
fixed oxide lattice, where certain surface oxygen atoms just take part in the 
reaction and the resulting vacancies are refilled afterwards,
a distinction between LH and MvK is in fact just semantics: One could equally well
view the oxide lattice with the created vacancies (thus MvK) as the real substrate, 
where then oxygen adsorbs into these sites and reacts thereafter (thus LH).
Although such a scenario has been advocated as a MvK mechanism for CO oxidation
on RuO${}_2$(110) \cite{over00}, one might even argue that it contradicts the 
original MvK definition in its strictest sense, which emphasized the non-finite
number of active sites (that is certainly fixed by the number of surface oxygen
sites in the latter situation). 

Other interpretations of MvK have therefore focused
more on a diffusional aspect that would arise if oxygen is consumed at one
point of the lattice surface but replenished elsewhere, requiring some form
of oxygen transport in or at the surface (that might even become rate-limiting and 
thus yield a new type of kinetics) \cite{knoezinger}. Nevertheless, a clearcut
distinction could even then not always be made. Picturing in particular a 
possibly only one-layer thick oxide
film, that does not necessarily resemble a bulk-like oxide structure at all,
or the aforedescribed situation of a surface close to an instability with
a continuous formation and destruction of a thin surface oxide (possibly
creating mesoscopic roughness due to the etching of metal particles from
the oxide framework \cite{hendriksen03b,hendriksen02}), how would one
be able to distinguish what is an oxide vacancy, what the oxide, what an
oxygen atom diffusing on or below the metal surface, let alone determine whether this
would show up in a unique kinetics that would still be describable by simple
macroscopic rate equations?

In this regard, another intriguing result of our thermodynamic considerations 
is the different stability range of surface oxides on Pd(111) and Pd(100).
Again, provided the catalyst is close to thermodynamic equilibrium, this
would imply that under certain $(T,p)$-conditions the active phase for
CO oxidation could be on-surface adphases (i.e. LH) on some facets of nanoparticles
and surface oxides (let's call this MvK) on others. A situation that 
could make the overall kinetic data (and its modelling) much more complex 
than hitherto assumed. In this particular sense, an understanding of the data 
from Pd could be much more difficult than say for Ag, where we would not expect
neither bulk nor surface oxides to play a significant role in CO oxidation
catalysis on the basis of the present results, and in agreement with the 
conclusions of a more detailed recent theoretical study \cite{li03}.

\section{Summary}

In conclusion we have derived simple thermodynamic stability conditions
for bulk and surface oxides in contact with an environment formed of CO
and O${}_2$. Applying these to the late $4d$ transition metal series
from Ru to Ag, we obtain a progressively smaller stability range for
the bulk oxides that extends well into the environmental conditions
representative of technological CO oxidation catalysis only in the case of 
RuO${}_2$. In particular for Pd, thin surface oxides could nevertheless still 
be stable in the more oxygen-rich part of this $(T,p)$-range, while 
for CO containing gas phases neither surface nor bulk oxide will be stable 
on Ag. Most intriguingly, oscillations between metal and bulk (surface) oxide 
phase for Rh (Pd) are conceivable under technologically relevant 
$(T,p)$-conditions. 

Provided the catalyst is close to the constrained
thermodynamic equilibrium forming the basis of our considerations, these
results may give first insight into the role of oxide formation played
in technological CO oxidation catalysis at these $4d$ TMs. A comprehensive
understanding of the microscopic state and full catalytic cycle must
however be based on statistical mechanics as the next step, i.e. by modelling
the very dynamic behavior and interplay of the manyfold of elementary
processes.

\section*{Acknowledgments}

We acknowledge detailed discussions with Cathy Stampfl and Weixue Li 
on the Ag/Ag${}_2$O system and with Mira Todorova on the Pd/PdO system.
K.R. is grateful for financial support from the Deutsche Forschungsgemeinschaft
(DFG). We thank Mark Miller for a careful reading of the manuscript.


\begin{references}

\bibitem{taylor93}
K.C. Taylor, Catal. Rev.-Sci. Eng. {\bf 35}, 457 (1993).

\bibitem{oh83}
S.H. Oh and J.E. Carpenter, J. Catal. {\bf 80}, 472 (1983).

\bibitem{kellog85}
G.L. Kellog, Phys. Rev. Lett. {\bf 54}, 82 (1985); Surf. Sci. {\bf 171}, 359 (1986).

\bibitem{peden88}
C.H.F. Peden, D.W. Goodman, D.S. Blair, P.J. Berlowitz, G.B. Fisher, and
S.H. Oh, J. Phys. Chem. {\bf 92}, 1563 (1988).

\bibitem{turner81}
J.E. Turner, B.C. Sales, and M.B. Maple, Surf. Sci. {\bf 103}, 54 (1981);
{\em ibid.} {\bf 109}, 591 (1981).

\bibitem{bondzie96}
V.A. Bondzie, P. Kleban, and D. J. Dwyer, Surf. Sci. {\bf 347}, 319 (1996). 

\bibitem{ziauddin97}
M. Ziauddin, G. Veser, and L. D. Schmidt, Catal. Lett. {\bf 46}, 159 (1997). 

\bibitem{veser99}
G. Veser, A. Wright, and R. Caretta, Catal. Lett. {\bf 58}, 199 (1999).

\bibitem{stampfl02}
C. Stampfl, M.V. Ganduglia-Pirovano, K. Reuter, and M. Scheffler,
Surf. Sci. {\bf 500}, 368 (2002).

\bibitem{over00}
H. Over, Y.D. Kim, A.P. Seitsonen, S. Wendt, E. Lundgren, M. Schmid,
P. Varga, A. Morgante, and G. Ertl, Science {\bf 287}, 1474 (2000).

\bibitem{kim01b}
Y.D. Kim, H. Over, G. Krabbes, and G. Ertl, Topics in Catalysis {\bf 14},
95 (2001).

\bibitem{hendriksen03a}
B.L.M. Hendriksen, S.C. Bobaru, and J.W.M. Frenken, Surf. Sci. ({\em submitted}).

\bibitem{hendriksen03b}
B.L.M. Hendriksen, {\em Model Catalysts in Action: High-Pressure Scanning
Tunneling Microscopy}, Ph.D. thesis, Universiteit Leiden (2003).

\bibitem{carlisle00b}
C.I. Carlisle, D.A. King, M.L. Boucquet, J. Cerd{\'a}, and P. Sautet, Phys. Rev. Lett.
{\bf 84}, 3899 (2000).

\bibitem{li03}
W.X. Li, C. Stampfl, and M. Scheffler, Phys. Rev. B {\bf 67}, 045408 (2003);
and to be published.

\bibitem{weinert86}
C.M. Weinert and M. Scheffler, In: {\em Defects in Semiconductors},
H.J. von Bardeleben (Ed.), Mat. Sci. Forum {\bf 10-12}, 25 (1986).
http://www.fhi-berlin.mpg.de/th/publications/Mat-Sci-Forum-10to12-25-1986.pdf .

\bibitem{scheffler88}
M. Scheffler, In: {\em Physics of Solid Surfaces - 1987}, J. Koukal (Ed.),
Elsevier, Amsterdam (1988).
http://www.fhi-berlin.mpg.de/th/publications/Phys-of-Sol-Surfaces-1987-1988.pdf;
M. Scheffler and J. Dabrowski, Phil. Mag. A {\bf 58}, 107 (1988).
http://www.fhi-berlin.mpg.de/th/publications/Phil-Mag-A-58-107-1988.pdf . 

\bibitem{kaxiras87}
E. Kaxiras, Y. Bar-Yam, J.D. Joannopoulos, and K.C. Pandey,
Phys. Rev. B {\bf 35}, 9625 (1987).

\bibitem{qian88}
G.-X. Qian, R.M. Martin, and D.J. Chadi, Phys. Rev. B {\bf 38}, 7649 (1988).

\bibitem{wang98}
X.-G. Wang, W. Weiss, Sh.K. Shaikhutdinov, M. Ritter, M. Petersen, F. Wagner,
R. Schl\"ogl, and M. Scheffler, Phys. Rev. Lett. {\bf 81}, 1038 (1998).

\bibitem{wang00}
X.-G. Wang, A. Chaka, and M. Scheffler, Phys. Rev. Lett. {\bf 84}, 3650 (2000).

\bibitem{reuter02a}
K. Reuter and M. Scheffler, Phys. Rev. B {\bf 65}, 035406 (2002).

\bibitem{reuter03a}
K. Reuter and M. Scheffler, Phys. Rev. Lett. {\bf 90}, 046103 (2003).

\bibitem{reuter03b}
K. Reuter and M. Scheffler, Phys. Rev. B ({\em submitted}); cond-mat/0301602.

\bibitem{JANAF}
D.R. Stull and H. Prophet, {\em  JANAF Thermochemical Tables}, 2nd ed.,
U.S. National Bureau of Standards, Washington, D.C. (1971).

\bibitem{reuter03c}
K. Reuter and M. Scheffler, (to be published).

\bibitem{CRC}
{\rm CRC Handbook of Chemistry and Physics}, 76th ed. (CRC Press, Boca Raton,
FL 1995).

\bibitem{grillo03}
M.E. Grillo, M.V. Ganduglia-Pirovano, and M. Scheffler, ({\em unpublished results}).

\bibitem{rogal03}
J. Rogal, M. Todorova, K. Reuter, and M. Scheffler ({\em to be published}).

\bibitem{lundgren02}
E. Lundgren, G. Kresse, C. Klein, M. Borg, J.N. Andersen, M. De Santis, Y. Gauthier,
C. Konvicka, M. Schmid, and P. Varga, Phys. Rev. Lett. {\bf 88}, 246103 (2002).

\bibitem{todorova03}
M. Todorova, E. Lundgren, V. Blum, A. Mikkelsen, S. Gray, J. Gustafson, M. Borg,
J. Rogal, K. Reuter, J.N. Andersen, and M. Scheffler, Surf. Sci. ({\em in preparation}).

\bibitem{todorova03b}
M. Todorova, K. Reuter, and M. Scheffler ({\em to be published}).

\bibitem{kolthoff96}
D. Kolthoff, D. J\"urgens, C. Schwennicke, and H. Pfn\"ur, Surf. Sci. {\bf 365},
374 (1996).

\bibitem{zheng02}
G. Zheng and E.I. Altman, Surf. Sci. {\bf 504}, 253 (2002).

\bibitem{szanyi93}
J. Szanyi, W.K. Kuhn, and D.W. Goodman, J. Vac. Sci. Techn. A {\bf 11(4)}, 1969 (1993).

\bibitem{reuter02b}
K. Reuter, C. Stampfl, M.V. Ganduglia-Pirovano, and M. Scheffler, Chem. Phys. Lett.
{\bf 352}, 311 (2002).

\bibitem{reuter02c}
K. Reuter, M.V. Ganduglia-Pirovano, C. Stampfl, and M. Scheffler, Phys. Rev. B {\bf 65}, 
165403 (2002).

\bibitem{hendriksen03c}
B.L.M. Hendriksen, M.D. Ackermann, and J.W.M. Frenken, 
Surf. Sci. ({\em submitted}).

\bibitem{masel96}
R.I. Masel, {\em Principles of Adsorption and Reaction on Solid Surfaces},
Wiley, New York (1996).

\bibitem{mars54}
P. Mars and D.W. van Krevelen, Chem. Eng. Sci. {\bf 3}, 41 (1954).

\bibitem{knoezinger}
G. Ertl, H. Kn\"ozinger, and J. Weitkamp (Eds.), {\em Handbook of
Heterogeneous Catalysis}, Wiley, New York (1997).

\bibitem{hendriksen02}
B.L.M. Hendriksen und J.W.M. Frenken, Phys. Rev. Lett. {\bf 89}, 
046101 (2002).


\end{references}
\end{document}